# Chiral Fermions on the Lattice

Rajamani Narayanan and Herbert Neuberger

Department of Physics and Astronomy

Rutgers University, Piscataway, NJ 08855-0849

## Abstract

An expression for the lattice effective action induced by chiral fermions in any even dimensions in terms of an overlap of two states is shown to have promising properties in two dimensions: The correct abelian anomaly is reproduced and instantons are suppressed.

The fundamental building blocks of matter, as known today, are chiral. If there are no anomalies, relativistic field theory can consistently describe chiral matter interacting via gauge boson exchange to any order in the Feynman diagram expansion. If matter were vector–like and all couplings were asymptotically free, work based on lattice formulations shows that one can sum all diagrams unambiguously. On the other hand, difficulties encountered in regularizing chiral fermions (even in asymptotically free situations) have raised doubts about the existence of chiral gauge theories at the non–perturbative level. This is a central problem of particle physics oriented field theory and its resolution may open up new possibilities in the endeavor to unify more of the observed forces in nature.

We focus on the lattice formalism because one can then regularize the bosonic degrees of freedom in a gauge invariant manner. The difficulties arising upon the addition of chiral lattice fermions are well known [1]. If we do not break gauge invariance in the fermionic sector, uninvited "doublers" are produced making the particle content vector–like and protecting the theory from ever having any anomalies. If we break gauge invariance, recovering it in the continuum limit for anomaly free representations is, at best, a matter of delicate fine tuning that would be very difficult to implement [2].

The modern approach [3] to the quantization of chiral gauge theories is a two step process: (a) Regularization of the chiral fermionic determinant in an arbitrary smooth gauge background. (b) Subsequent integration over all gauge fields. If there are no anomalies, the answer to the first step is gauge invariant and the subsequent functional integration over the gauge fields can be attempted. The problem of regularizing chiral fermions on the lattice is encountered at the first step and we concern ourselves mainly with this step in this letter. However, our formulation accepts arbitrary lattice gauge backgrounds. Some readers may object to the strategy which led us to the two steps in the first place, because fermionic and bosonic fields are quantized sequentially rather than simultaneously. We cannot overrule this objection; however, the progress in our understanding of anomalies in the mid–eighties [3] and the discovery of the important instanton effects in the seventies [4] were made within this two-step framework.

An easy way of presenting a new approach due to Kaplan [5] is as follows [6]. We start with a generic vector–like gauge theory that is easy to regularize.

$$\mathcal{L} = \mathcal{L}_1 + \mathcal{L}_g(A), \quad \mathcal{L}_1 = i\bar{\psi}(\slashed{\partial} - ig\slashed{A})\psi + \bar{\psi}(MP_+ + M^\dagger P_-)\psi \qquad (1)$$

$P_\pm = \frac{1\pm\gamma_5}{2}$ and $A_\mu = A_\mu^a T^a$ where $T^a$'s are generators in the appropriate representation of the gauge group. $\mathcal{L}_g(A)$ is the pure gauge part, whose explicit form is irrelevant to us here. To achieve chiral symmetries, we need massless fermions. $M$ is a square matrix and therefore $M$ and $M^\dagger$ will have equal rank implying that massless fermions will occur in pairs of opposite chirality. We can overcome this only if we make $M$ infinite dimensional and endow $M$ with a nonzero analytical index. The infinite dimensional "flavor" space, whose sole role is to deal with the specific problem of chiral massless fermions, adds an additional source of indeterminacy to our problem. We can therefore regularize the ultraviolet infinities in any way we wish (since they are "released" from the "duty" to also induce anomalies), and we choose the lattice. The specific problems having to do with chirality are relegated to the step when we try to control the infinity of the internal (flavor) space.

The simplest way to realize a non–trivial index for $M$ is to choose the internal space as the real line and choose $M = \partial_s + f(s)$ with $f(s)$ asymptotically approaching constants of different signs when $s \to \pm\infty$. This produces the wall system discussed by Callan and Harvey [7]. $s$-space can also be discretized and the entire Callan–Harvey analysis can be transferred to the lattice. This observation is due to Kaplan [5] who also showed that on the lattice one may work with a defect strictly local in $s$.

We deal with the additional infinity by viewing internal space as the Euclidean time axis of an $s$–dependant Hamiltonian $\hat{H}$ that is subjected to a mass shock at $s = 0$ [8]. The chiral determinant is replaced by the overlap of two ground states, corresponding to $\hat{H}_\pm$ for $s > 0$ and $s < 0$ respectively. A straightforward derivation [8] gives

$$\hat{H}_\pm = \hat{a}^\dagger \mathbf{H}_\pm \hat{a} \quad e^{\mathbf{H}_\pm} = \begin{pmatrix} \frac{1}{\mathbf{B}^\pm} & \frac{1}{\mathbf{B}^\pm}\mathbf{C} \\ \mathbf{C}^\dagger \frac{1}{\mathbf{B}^\pm} & \mathbf{C}^\dagger \frac{1}{\mathbf{B}^\pm}\mathbf{C} + \mathbf{B}^\pm \end{pmatrix} \qquad (2)$$

$\hat{a}_{nAi}$ and $\hat{a}^{\dagger}_{nAi}$ are fermion operators satisfying canonical anticommutation relations: $\{\hat{a}_{nAi},\hat{a}^{\dagger}_{mBj}\} = \delta_{nm}\delta_{AB}\delta_{ij}$, $\{\hat{a}_{nAi},\hat{a}_{mBj}\} = 0$, $\{\hat{a}^{\dagger}_{nAi},\hat{a}^{\dagger}_{mBj}\} = 0$. The indices $A$ and $B$ run over $2^{d/2}$ spinor components and $i$ and $j$ over group indices. $n$ is the d-dimensional space index. $\mathbf{B}^{\pm}$ and $\mathbf{C}$ are matrices depending on the gauge fields. Explicitly,

$$\mathbf{B}^{\pm}_{n\alpha i, m\beta j} = (d+1 \mp m)\delta_{nm}\delta_{\alpha\beta}\delta_{ij} - \frac{1}{2}\delta_{\alpha\beta}\sum_{\mu}\left[\delta_{m,n+\hat{\mu}}U^{ij}_{n,\mu} + \delta_{n,m+\hat{\mu}}U^{ji}_{m,\mu}{}^{*}\right]$$

$$\mathbf{C}_{n\alpha i, m\beta j} = \frac{1}{2}\sum_{\mu}\left[\delta_{m,n+\hat{\mu}}U^{ij}_{n,\mu} - \delta_{n,m+\hat{\mu}}U^{ji}_{m,\mu}{}^{*}\right]\sigma^{\alpha\beta}_{\mu} \quad (3)$$

$U_{n,\mu}$ is a matrix in the appropriate representation of the gauge group and is associated with the gauge field on the link at site $n$ in the postive $\mu$ direction. We have used the following representation for the $\gamma$ matrices: $\gamma_{\mu} = \begin{pmatrix} 0 & \sigma_{\mu} \\ \sigma^{\dagger}_{\mu} & 0 \end{pmatrix}$ $\gamma_{5} = \begin{pmatrix} 1 & 0 \\ 0 & -1 \end{pmatrix}$, where $\sigma_{0} = i$ and $\sigma_{j}; j = 1,..,d-1$ are the $2^{d/2-1} \times 2^{d/2-1}$ generalization of Pauli matrices to $d$ (even) dimensions. $0 < m < 1$ is a mass parameter. The two Hamiltonians in (2) are hermitian due to the hermiticity of (3). One can compute the determinants of $e^{\mathbf{H}_{\pm}}$ and prove that the $\mathbf{H}_{\pm}$ are traceless [8]. While this indicates the presence of both positive and negative eigenvalues the relation between the two sets of eigenvalues and eigenvectors is not as simple as it would be in the continuum.

The gauge action induced by the integration over fermions is then given by the following formula:

$$e^{S_i(U)} = \frac{{}_U\langle 0-|0+\rangle_U}{{}_1\langle 0-|0+\rangle_1}e^{i[\Phi_+(U)-\Phi_-(U)]} \quad (4)$$

The $|0\pm\rangle_U$ are ground states of $\hat{H}_{\pm}(U)$ and the phases $\Phi_{\pm}(U)$ are defined by a convention borrowed from Wigner–Brillouin perturbation theory:

$$e^{i\Phi_{\pm}(U)} = \frac{{}_U\langle 0\pm|0\pm\rangle_1}{|{}_U\langle 0\pm|0\pm\rangle_1|} \quad (5)$$

(4) completely defines the real part of $S_i(U)$. (5) is one choice to fix the phase ambiguity in (4). It is only in (5) that gauge invariance is broken; any other definition of $\Phi_{\pm}(U)$ which differs from the above by a local functional of $U - 1$ (i.e. when expanded in $U - 1$ the coefficients are analytic at the origin of lattice momentum space) will be acceptable. The real part of $S_i$ is naturally gauge invariant; another way to enforce the gauge invariance of $Re(S_i)$ is proposed in [9].

A continuum variant of the overlap formula was shown recently to give the correct anomaly in two Euclidean dimensions with a $U(1)$ gauge group [8]. The calculation is tedious because there are no straightforward Feynman–diagram techniques and one has to use some form of Hamiltonian perturbation theory. To do the same calculation on the lattice would be even more tedious due to the extra nonlinearity in the expressions for the transfer matrices. In addition, such a calculation would not tell us whether nonperturbatively $S_i$ behaves correctly at the semiclassical level. Therefore we proceed numerically.

We set $d = 2$ and work on a square lattice of size $L \times L$. We use antiperiodic boundary conditions for the fermions to avoid divergences associated with constant spinors. For a given gauge configuration the matrices $e^{\mathbf{H}_\pm}$ are constructed in momentum basis and diagonalized by the Jacobi method. The corresponding one particle states are filled to construct the vacua entering the overlaps. The overlaps in (4) and (5) are evaluated by computing the appropriate determinants.

For sufficiently small $|U - 1|$ the gauge fields are perturbative and each of the $\mathbf{H}_\pm$ have $L^2$ positive eigenvalues. The determinants are non-vanishing and $S_i(U)$ in (4) is obtained. To check the anomaly we choose $U_{n,\mu}^{(k)} = \exp[i\frac{A_\mu(\phi)}{L}\cos(\frac{2\pi n \cdot k}{L} + \frac{\pi k_\mu}{L})]$ as the gauge configuration on the lattice. $k_\mu \in Z$ is kept fixed as $L \to \infty$. $\{U_{n,\mu}^{(k)}\}$ represents a standing plane wave with a fixed number of nodes on the torus in the continuum limit. $A_\mu(\phi) = A_\mu + 2\phi \sin \frac{\pi k_\mu}{L}$ represents a family of gauge equivalent connections. When $\phi$ is varied, the imaginary part of $S_i(U)$ should change and the change in the continuum limit is

$$\lim_{L \to \infty} -\frac{i}{\pi}\frac{\partial S(A_\mu(\phi))}{\partial \phi}\bigg|_{\phi=0} = \frac{1}{4\pi}[A_1 k_2 - A_2 k_1]. \tag{6}$$

We fix $k = (1, 0)$, $A_1 = A_2 = 0.32$ and, in preparation for $L \to \infty$ limit, compute the LHS in (6) as a function of $L$. The computations are performed at two different values of $m$.

The data points are shown in Fig.1 as a function of $1/L^2$. As expected, the continuum limit is independent of $m$. Extrapolation to $L = \infty$ is smooth and yields $-0.02545(5)$ for the anomaly which agrees with the continuum value of $-0.02546$.

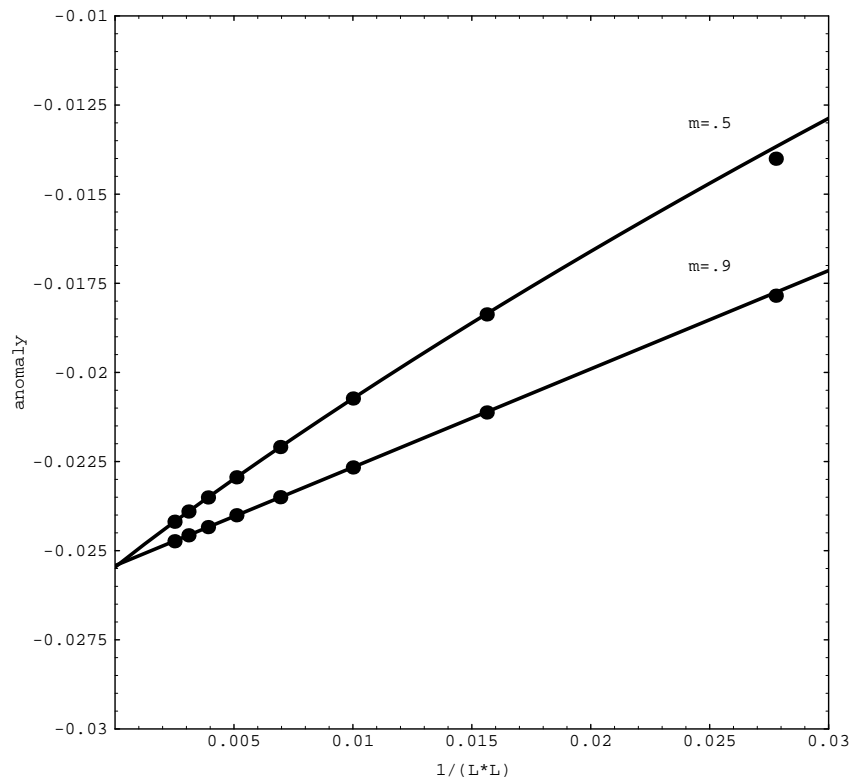

**Figure 1.** The anomaly (LHS of (6)) as a function of lattice size. The points are for $L = 6$ to $20$ in steps of $2$ at $m = 0.5, 0.9$. The line are fits to the data for $L \geq 10$.

Two comments are in order here. Since we used plane waves, once several $k$'s are checked, the numerical work is as good a check as an analytical calculation would be. The coefficient in front of the curvature in the anomaly equation is $\frac{1}{4\pi}$ and not $\frac{1}{2\pi}$ as obtained by Jansen [11] in the three dimensional wall set-up. This reflects the fact that the current considered by Callan and Harvey is afflicted by the covariant form of the anomaly not the consistent one. Since we deal with the effective action directly we have to obtain the "consistent" value [8, 12].

The overlap in (4) will vanish when the numbers of positive eigenvalues for $\mathbf{H}_\pm$ are different. This situation is expected if the set of gauge fields represent a U(1) connection on a non-trivial principle bundle over the two-dimensional torus. The first Chern number, represented by the "lattice topological charge",

$$C_1 = \sum_\diamond \frac{\log(U_\diamond)}{2\pi i} \qquad (7)$$

should be non-zero in such an instance [13]. In (7), $\log(1) = 0$, the cut is along the negative real axis, and the sum runs over all elementary palquettes. $U_\diamond$ is the parallel transporter around the plaquette $\diamond$. A simple choice that produces a non-zero $C_1$ is $U_\diamond = \exp[\frac{2\pi i q}{L^2}]$ for all plaquettes with $q$ some fixed integer as $L \to \infty$. $C_1 = q$ and this configuration has a smooth continuum limit. A gauge configuration that produces this uniform $U_\diamond$ is $U_{(n_1,n_2),1} = \exp[\frac{2\pi i q}{L^2} n_2]$, $U_{(n_1,L-1),2} = \exp[\frac{2\pi i}{L} n_1]$ and all other $U_{n,2}$ are equal to unity. For this gauge configuration at $L = 6$, we found that $\mathbf{H}_+$ had $36 - q$ positive eigenvalues and $\mathbf{H}_-$ had 36 positive eigenvalues for all $|q| \leq 5$. That only $\mathbf{H}_+$ is affected and not $\mathbf{H}_-$ is consistent with the observation by Golterman et. al. [10] that the Goldstone-Wilczek currents are present only on one side of the defect. We added random noise to the above gauge configuration and found that the qualitative behavior was robust. Generically, the behavior we obtain is clearly of the kind we would expect if instanton effects are to be reproduced. This was a problem for many other approaches to the problem of regularizing chiral gauge theories as emphasized in [14]. It would be interesting to insert the appropriate fermion operators to obtain a non-zero result and carry over the 't Hooft computation [4] of fermion number violating processes to the lattice.

The extension of the results in this paper to four dimensions is just a question of computer time. We think it would be feasible within current computer capacities. In two dimensions, for modest volumes, it appears feasible to proceed to the next stage, i.e. estimate the average over gauge fields. This by itself might be quite interesting because even in two dimensions chiral theories are either exactly soluble or, if also strongly coupled, quite intractable. One may also revisit attempts to construct anomalous gauge theories: for example one may check whether the $a = 1$ case of the Jackiw and Rajaraman [15] continuum construction can be reproduced on the lattice.

We should stress that we have invested little time to date in thinking about efficient methods of computation in four dimensions. We were only concerned with what one could do if computer power were unlimited. We believe that the approach outlined in this paper has sufficient potential to warrant more intense scrutiny.

## Acknowledgements

We would like to thank T. Banks, D. B. Kaplan, M. F. Golterman, S. Shenker and E. Witten for discussions. E. Witten was probably aware of a connection between the chiral determinant and some form of overlap since the mid-eighties. This research was supported in part by the DOE under grant # DE-FG05-90ER40559.